\documentclass[12pt,aasmacros]{spieman}  
\usepackage{amsmath,amsfonts,amssymb}
\usepackage{color,soul}
\usepackage{graphicx}
\usepackage{setspace}
\usepackage{tocloft}
\usepackage{subfig}
\usepackage{tabularx}

\title{Imaging effects due to pixel distortions in CdZnTe (CZT) detectors - results from the HREXI Calibration Facility (HCF).}

\author[a, b, *]{Arkadip Basak}
\author[a]{Branden Allen}
\author[a]{Jaesub Hong}
\author[a]{Daniel P. Violette}
\author[a]{Jonathan Grindlay}
\affil[a]{Harvard University - Center for Astrophysics, 60 Garden Street, Cambridge, MA - 02138, USA.}
\affil[b]{Anton Pannekoek Institute, University of Amsterdam, Science Park - 904, 1098XH Amsterdam, NL.}

\cftpagenumbersoff{figure}
\cftpagenumbersoff{table} 
\begin{document} 
\maketitle

\begin{abstract}
ProtoEXIST2 (P2) was a prototype imaging X-ray detector plane developed for wide-field Time Domain Astrophysics
(TDA) in the 5 - 200 keV energy band.  It was composed of an 8~$\times$~8 array of 5~mm thick,
2~cm~$\times$~2~cm pixelated (32 $\times$ 32) CdZnTe (CZT) detectors with a 0.6~mm pitch that utilize the NuSTAR ASIC
(NuASIC) for readout \cite{nustar, Hong2013_2, spie11}.  During the initial detector development process leading up to post-flight examination of the entire detector plane, distortions in expected pixel positions and
shapes were observed in a significant fraction of the detectors \cite{spie11,subpix}.
The HREXI (High Resolution Energetic X-ray Imager) Calibration Facility (HCF) was designed and commissioned to improve upon these early
experiments and to rapidly map out and characterize pixel non-uniformities and defects within CZT detector planes at resolutions down to 50~$\rm \mu$m.  Using this facility, the sub-pixel level detector response of P2 was measured at 100~$\rm 
 \mu$m\cite{basak_2020}
resolution and analyzed to extract and evaluate the area and profile of individual pixels, their
morphology across the entire P2 detector plane for comparison with previous measurements and to
provide additional characterization\cite{subpix}. In this article, we evaluate the imaging 
performance of a coded-aperture telescope using the observed pixel morphology for P2 detectors.  This investigation will serve as an
initial guide for detector selection in the development of HREXI detector planes, for the future implementation of the 4pi X-Ray Imaging Observatory
(4piXIO)\cite{josh2019} mission which aims to provide simultaneous and continuous  imaging of the full sky ($\rm 4\pi$ sr) in the 3-200 keV energy band with $\rm \simeq$ 2 arcmin angular resolution and $\simeq$ 10 arcsec source localization, as well as other, future coded-aperture instruments.

\end{abstract}


{\noindent \footnotesize\textbf{*}Arkadip Basak,  \linkable{arkadip.basak@cfa.harvard.edu} }

\begin{spacing}{1}   

\section{Introduction}
\label{intro}

The High-Resolution Energetic X-Ray Imager (HREXI) program was initiated for the development of a
modular and highly scalable CdZnTe (CZT) detector plane architecture for broadband, high resolution (both spatial and spectral)  space-borne wide-field coded aperture X-ray telescope. HREXI will enable new investigations for high energy
time-domain astrophysics (TDA) to probe the properties and populations of all classes of 
X-ray (and soft gamma-ray) transients arising from accretion onto black holes, neutron stars, and white dwarf X-ray binary populations as
well as enable high angular resolution searches for gravitational wave counterparts detected with
the next generation gravitational wave interferometers coming online over the next decade\cite{josh2019}.  

The HREXI (successor to ProtoEXIST2) detector plane architecture is a close-tiled array of pixelated CZT detectors, each directly bonded to the NuSTAR ASIC (NuASIC)\cite{nustar}.  A 32 $\times$ 32 array of
anode inputs (gold studs) on a linear X-Y pitch of 0.6048 mm are bonded on the upper surface of
the NuASIC\cite{nustar}.  These are each directly bonded with a conductive epoxy to a
matching array of square 550 $\mathrm{\mu}$m anode pixels deposited (Au) on the anode surface of Redlen
20 $\times$ 20 mm CZT crystals, each 3 mm thick\cite{Hong2013, Hong2013_2, spie11}.  The full ProtoEXIST2 (P2) detector plane was constructed from a closely-tiled 8 $\times$ 8 (numbered from 1 to 64) array of these Detector Crystal Units (DCUs) and demonstrated the first prototype large-scale tiling of  CZT/NuASICs. In P2 each individual CZT detector was 5 mm thick and biased with a nominal -600 V operating
voltage ($\rm V_{bias}$). More recent implementations utilize 3 mm thick CZT under a -400 $\rm V_{bias}$ (nominal).

P2 was flown as the active element of a high-altitude (39 km)\cite{hong2017} balloon-borne coded aperture telescope
and recovered in pre-flight condition during October of 2012, demonstrating stable operation in a
space-like environment.  Subsequently, P2 underwent extensive post-flight calibration and characterization
including a precise 1D scan with a collimated Am-241 source carried out independently on two axes
over the full detector plane.  This revealed the presence of substantial pixel-position and size
distortion in some detectors \cite{subpix}.  Further investigation and confirmation of this
effect were enabled with the completion of the HREXI Calibration facility (HCF)\cite{basak_2020},
which is capable of mapping the 2D response of individual pixels on a scale down to 50 $\mathrm{\mu}$m.  The causes of these defects are unknown and are under active investigation.  Since this is a persistent phenomenon observed in different batches of CZT grown a year apart over an 8-year span, a study was carried out in order to evaluate the potential consequences of these distortions on the imaging capabilities of
P2 and future space-borne wide-field instruments utilizing these detectors. Here, we 
evaluate the effects of distortions on the imaging capability using the P2 telescope parameters.

Coded Aperture Imaging is an imaging technique that has found widespread use in X-ray
astronomy since its inception nearly 4 decades\cite{caroli87} ago and continues to be an indispensable tool for TDA in the hard X-ray regime.  In general, this technique may be implemented in 1-D or 2-D using a
collimator with a known hole pattern and a position-sensitive detector which are mounted a fixed
distance from one another.  An X-ray source in the sky above the mask will posit photons on
the detector plane through the open mask elements resulting in a projection of the mask pattern in
X-rays on the surface of the detector plane which in turn is collated within detector data. We are then able to recover the position of the X-ray source through cross-correlation of the mask pattern with the reconstructed detector-plane image. In the case of the detector in our lab for P2, a 2D random pixel pattern with a
$\simeq$ 1.1mm linear pitch is utilized for the mask, where the detector plane is composed of a planar
pixelated detector with $\simeq$ 0.6~mm pitch located $\simeq$~880~mm below the mask surface. The mask and the detector planes are mounted parallel to one another.

This article has been organized in the following manner. Section \ref{pix-stack} describes our novel pixel-stacking technique, section \ref{reconstruction} describes a reconstructed stacked pixel after cross-correlating with a single mask element, section \ref{detector_anal} extends the same set of analyses to an ensemble of detectors, section \ref{correction} provides an elementary strategy to correct for pixel distortions while section \ref{conclusions} refers to primary results and future scope of work in the context of 4piXIO.


\section{HCF data and analysis}
\label{analysis}

\begin{figure}%
    \centering
    \subfloat{{\includegraphics[width=0.7\textwidth]{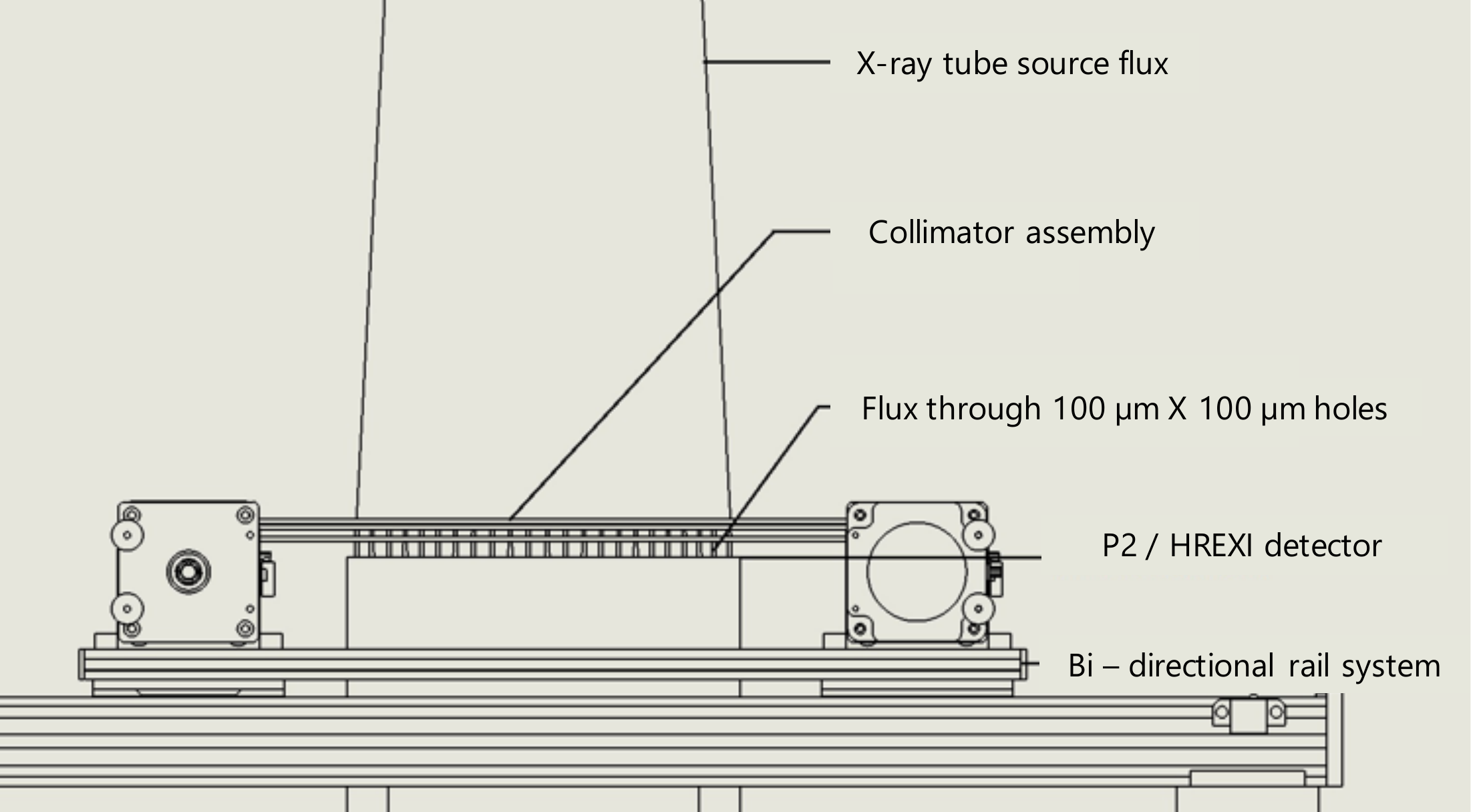} }}%
    \caption{Schematic showing the X-ray tube source flux, the collimator assembly, and simultaneous illumination of several 100~$\mu$m $\rm \times$ 100~$\mu$m pixels as either the P2 or the HREXI detector plane move along a bi-directional (X,Y) rail system. Figure adapted from Basak et. al. 2020\cite{basak_2020}}%
    \label{hcf-schematic}%
\end{figure}

\begin{figure}%
    \centering
    \subfloat{{\includegraphics[width=0.94\textwidth]{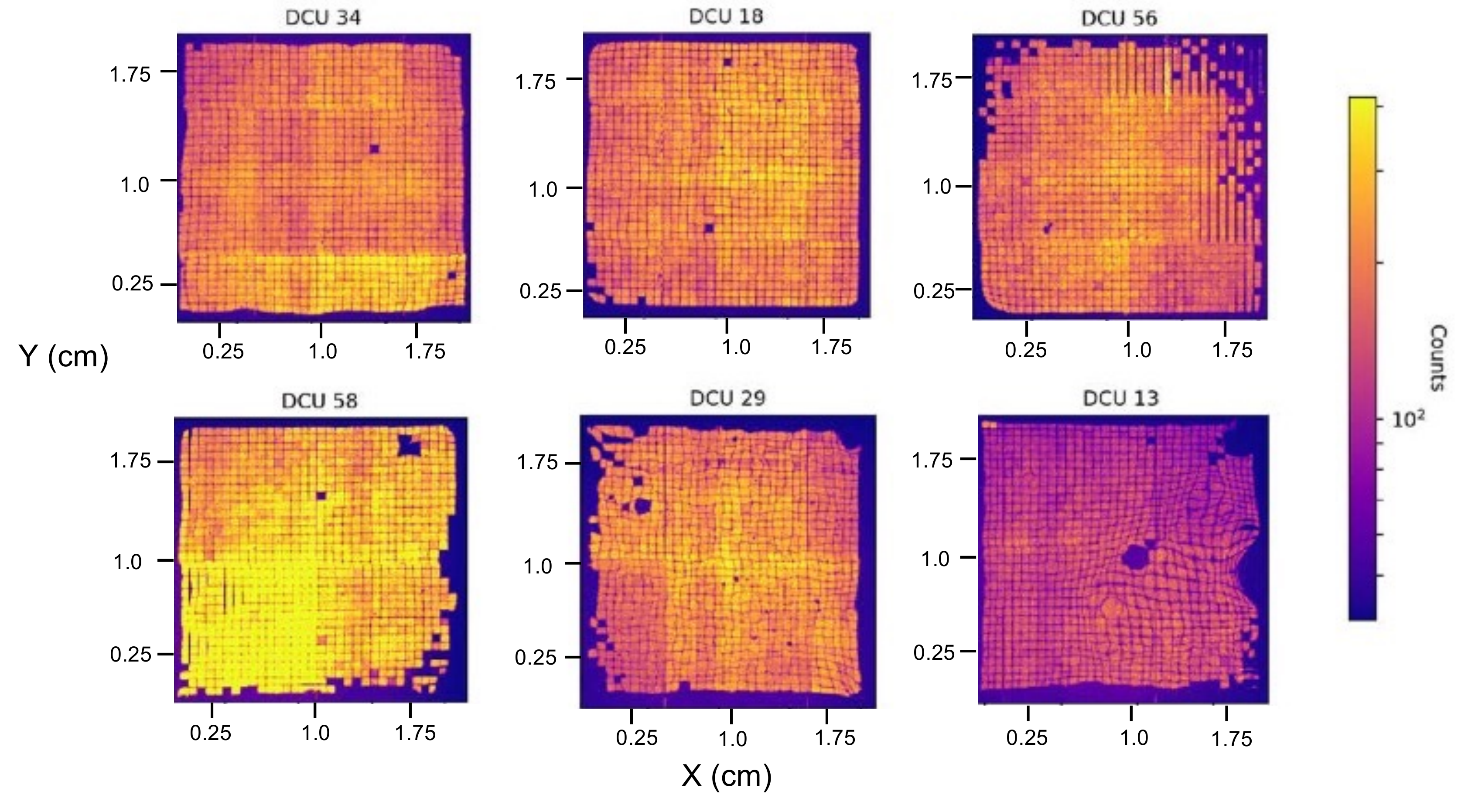} }}%
    \caption{HCF Detector maps of the P2 detector plane for DCUs 34, 18, 56, 58, 29, and 13. The size of each DCU is 2cm $\rm \times$ 2cm}%
    \label{detector_map}%
\end{figure}

The HREXI Calibration Facility (HCF) is comprised of an X-ray tube source\cite{MiniX} mounted $\simeq$ 2.2~m above a
tungsten collimator\cite{basak_2020}(mask) with an array 4 or 8\cite{basak_2020} etched 100~$\mu$m $\rm \times$ 100~$\mu$m square holes distributed on a 2~cm~$\times$~2~cm DCU grid. The design of the mask assembly and the orientation of the mask holes are constrained by the P2 detector event rate handling capability as mentioned explicitly in section 5 of Basak et. al. 2020\cite{basak_2020}. The collimator is mounted to a computer-controlled bi-directional (X $\times$ Y) linear rail system with an accuracy/step size of 5~$\mu$m.  Mapping of the response of individual pixels is carried
out by placing the detector(s) underneath the mask, turning on the X-ray tube source, and exposing the
detector to the collimated beam for approximately 10~s at each location. The mask array is then moved to the next X, Y position to be measured. Data collection continues
while the mask is moved in a boustrophedon pattern\cite{basak_2020} until the entire detector has been exposed to the
beam. Figure \ref{hcf-schematic} elucidates the X-ray source flux, the collimator, the detector(s), and the precision bi-directional rail system in HCF with a schematic.
The collected event data from the detector are combined with the collimator position
information to produce a response map for each of the individual pixels in the detector plane.
Co-addition of the response of all pixels for single-pixel events, i.e. events that contain nearly
all of the charge induced by an incident X-ray for each individual pixel, are co-added producing a full-detector response map
(e.g., Figure \ref{detector_map}). 

The HCF\cite{basak_2020} has been designed to enable rapid characterizations of these detector planes in
timescales $\simeq$ 3 days. However, since HREXI detectors are still under development as mentioned
in section \ref{intro}, this article presents results from calibration runs of its predecessor P2.
Individual P2 detectors, or detector crystal units (DCUs), consist of a single 5~mm thick CZT crystal with a 32~$\times$~32 anode pattern on a 604.8~$\rm \mu m$ pitch hybridized directly to a NuSTAR ASIC (NuASIC) that is used to readout individual events and their energies. Figure \ref{detector_map} presents a representative sample of DCU maps reconstructed from data collected with the HCF. 
The blank pixels on some of the DCUs have either been identified as hot and were disabled during the collection of data or, in some cases, the connection between the CZT crystal and the ASIC has failed, resulting in non-functional pixels\cite{basak_2020}. For annotating hot pixels, DCUs with large pileup effects were identified, followed by looking at anomalous trends in pixel counts in the unexposed portions of a DCU. These trends are usually characterized by a pixel having $\rm \geq 2 X$ counts when compared to an adjacent set of pixels. These large discrepancies can't be attributed to the changes in X-ray tube source\cite{MiniX, basak_2020}, the aberrations of mask hole patterns\cite{basak_2020}, or a difference in mask layer thicknesses\cite{basak_2020} and were classified as `hot'.

DCUs 34 and 13 represent two extreme cases within P2 that were detected through HCF-derived DCU maps; one detector displays significant
warping of the pixel pattern throughout the entire DCU, while the other shows uniform pixel response except at the corners. DCU 18 is comparable to DCU 34, except for a few other blank pixels on one of the corners. The DCU maps become incrementally warped and unconstrained within the pixel boundaries as one
traverses from DCU 34 to DCU 13 in figure \ref{detector_map}, making the analyses of DCUs 34 and 13 from section \ref{pix-stack} a representation of a set of extrema in the P2 detector plane. Note that the relative intensity differences seen in some pixels are the results of non-uniformities in the HCF collimator hole size which results in enhanced flux in quadrants of some of the individual detectors\cite{basak_2020} which has not been calibrated out here. However, they have been accounted for in the subsequent section(s). The figures in section \ref{pix-stack} to \ref{angres_sec} have also been limited to DCUs 34 and 13, while the P2 detector plane as a whole has been considered in section \ref{detector_anal}.

\subsection{Pixel Stacking}
\label{pix-stack}
After obtaining DCU maps, we have devised a `pixel stacking' technique that enables the simultaneous visualization and analysis of the morphologies of collections of any number of pixels within our detector plane, which is necessary due to the nature of our scanning and standard data-analysis techniques.  The time-resolved event data are used to derive a response map (event rate) for each of the individual pixels as a function of the mask position within the nominal $\simeq$ 1~cm~$\times$~1~cm scan region of a 2~cm~$\times$~2~cm DCU. The scan region of $\simeq$ 1~cm~$\times$~1~cm stems from using 4 holes per DCU in the mask assembly as mentioned specifically in sections 2.1 and 5 of Basak et. al. 2020\cite{basak_2020}. The scan regions for individual beams are then offset and combined to produce maps in physical space which yields the sub-pixel response maps presented here.  A small overlap in the scan region enables the direct comparison and removal of relative differences arising from hole-to-hole variation in our collimator across the full detector region as mentioned explicitly in Basak et. al. 2020\cite{basak_2020}. A stacked sub-pixel for $\rm n^{th}$ DCU in the P2 detector plane is defined as,

\begin{equation}
\label{pixstack}
    \rm P_{i,j|n} = \sum_{k = 1}^{1024} \frac{1}{G_k} C_{ i,j|k} \hspace{0.5em},
\end{equation}

Indices $\rm i,j$ denote each element in the sub-pixel space whereas index $\rm k$ represents each of the 1024 pixels in DCU n. $\rm P_{i,j|n}$ is the counts in a stacked sub-pixel $\rm (i,j)$, $\rm C_{i,j|k}$ represents the counts in each sub-pixel for pixel $\rm k$. The normalization factor $\rm G_k$ denotes the total counts in pixel $\rm k$ and hence takes care of inter-pixel count rate differences within a DCU due to fluctuations in Mini-X X-ray source\cite{MiniX} flux and/or differences in mask hole sizes. $\rm G_k$ is calculated by imposing a $\simeq$ 60-100 sub-pixels boundary for each pixel $\rm k$ depending on the DCU being considered, followed by a summation of their responses, only when pixel $\rm k$ is illuminated. This essentially means that the indices $\rm i,j$ run between 1 and 60-100. The size  of the sub-pixel boundary is limited to $\rm \simeq$ 60 sub-pixels for a DCU 34 while it is $\rm \simeq$ 100 sub-pixels for DCU 13.

\begin{figure}%
    \centering
    \subfloat{{\includegraphics[width=0.94\textwidth]{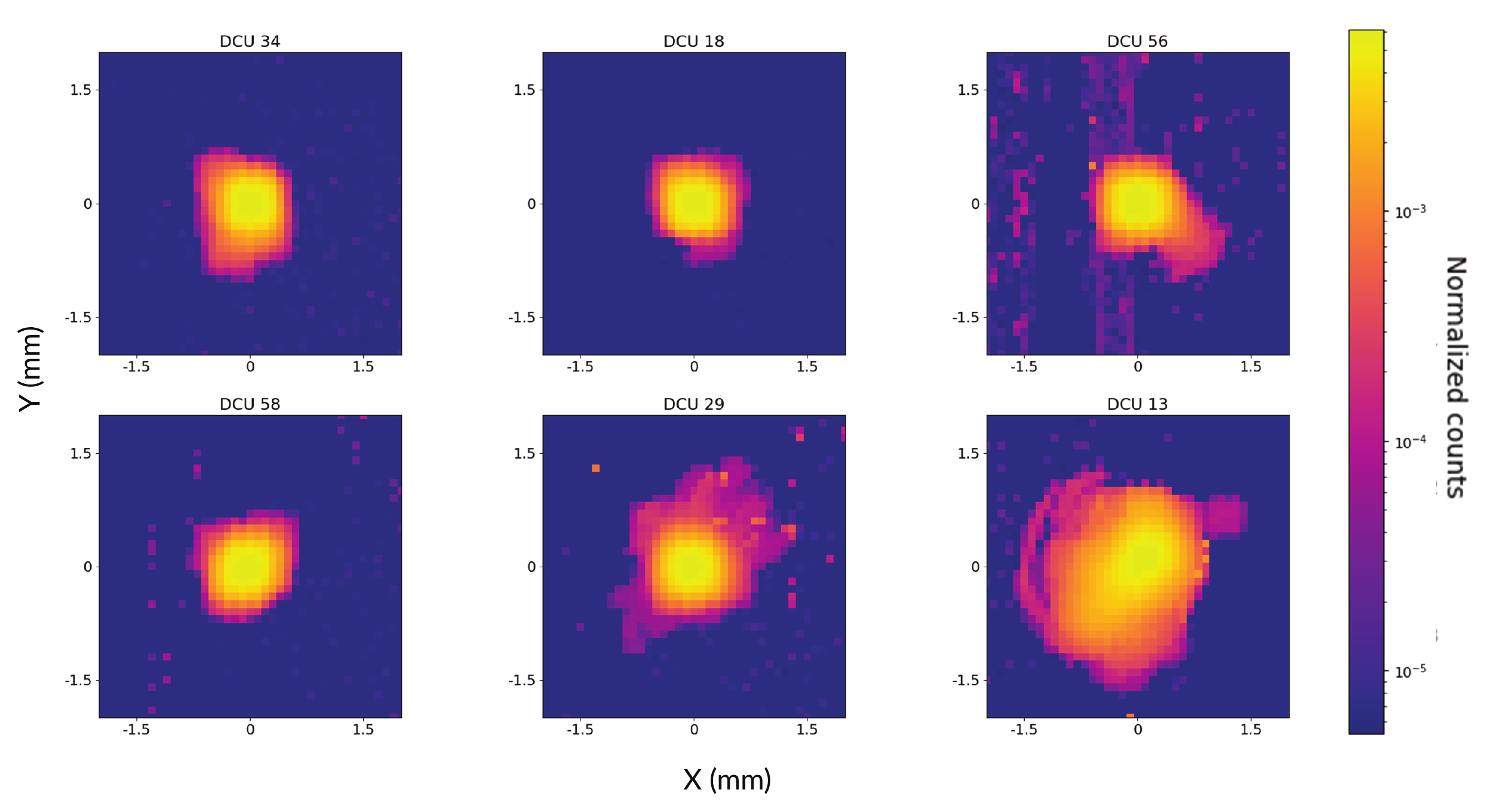} }}%
    \caption{Stacked pixel image of DCUs 34, 18, 56, 58, 29, and 13. The confines of a stacked pixel image increase progressively from DCUs 34 to 13. The stacked pixel representation of DCU 34 is well constrained within a $\simeq$ 1.5 mm$\times$ 1.5 mm sub-pixel grid, whereas the stacked pixel image of DCU 13 requires at least a $\simeq$ 2.0 mm $\times$ 2.0 mm grid to constrain itself. The coordinate (0,0) represents the true center of the stacked DCU pixel.}%
    \label{stack}%
\end{figure}

\begin{figure}%
    \centering
    \subfloat{{\includegraphics[width=0.98\textwidth]{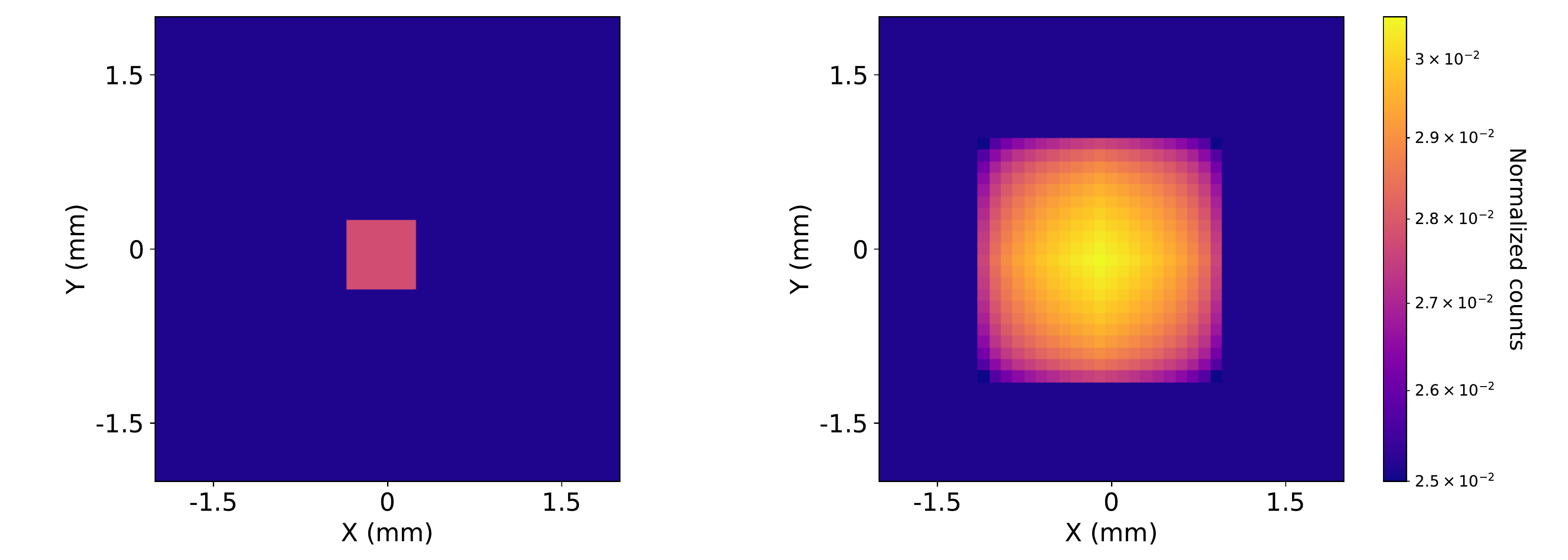}}}%
    \caption{A flux-normalized ideal P2 DCU stacked pixel(left) without any non-uniformities and a reconstructed stacked pixel image after cross-correlating the DCU stacked pixel with a mask element(right) size of 1.1 mm.}%
    \label{cc_ideal}%
\end{figure}

Figure \ref{stack} illustrates the stacked pixel plots for DCUs 34, 18, 56, 58, 29, and 13, while figure \ref{cc_ideal}(left) demonstrates an ideal P2 DCU `stacked pixel'. The flux in figure \ref{cc_ideal} has been normalized to conform with that of a nominal DCU pixel in HCF and to match the counts in figure \ref{stack}. The distribution of counts for the DCUs 34 and 18 are well constrained within a $\simeq$ 1.5 mm$\times$ 1.5 mm sub-pixel grid, whereas the presence of a significant number of warped pixels in DCUs 56, 58, 29, and 13 results in a diffused distribution of sub-pixel counts. The disabled pixels which appear as pixels with zero counts in the sub-pixel response maps have been left out of these analyses. A stacked DCU pixel essentially represents the entire DCU response superimposed on a single-pixel grid, with the inter-pixel flux differences either due to fluctuations in X-ray source flux and variations in mask hole sizes across a DCU or sections of DCUs corrected out. Cross-correlating a stacked DCU pixel with a single P2 mask element yields an image, encoding the entire DCU response when irradiated by an on-axis uniform source as mentioned in section \ref{intro}. 

\subsection{Stacked pixel Image reconstruction}
\label{reconstruction}
\begin{figure}%
    \centering
    \subfloat{{\includegraphics[width=0.94\textwidth]{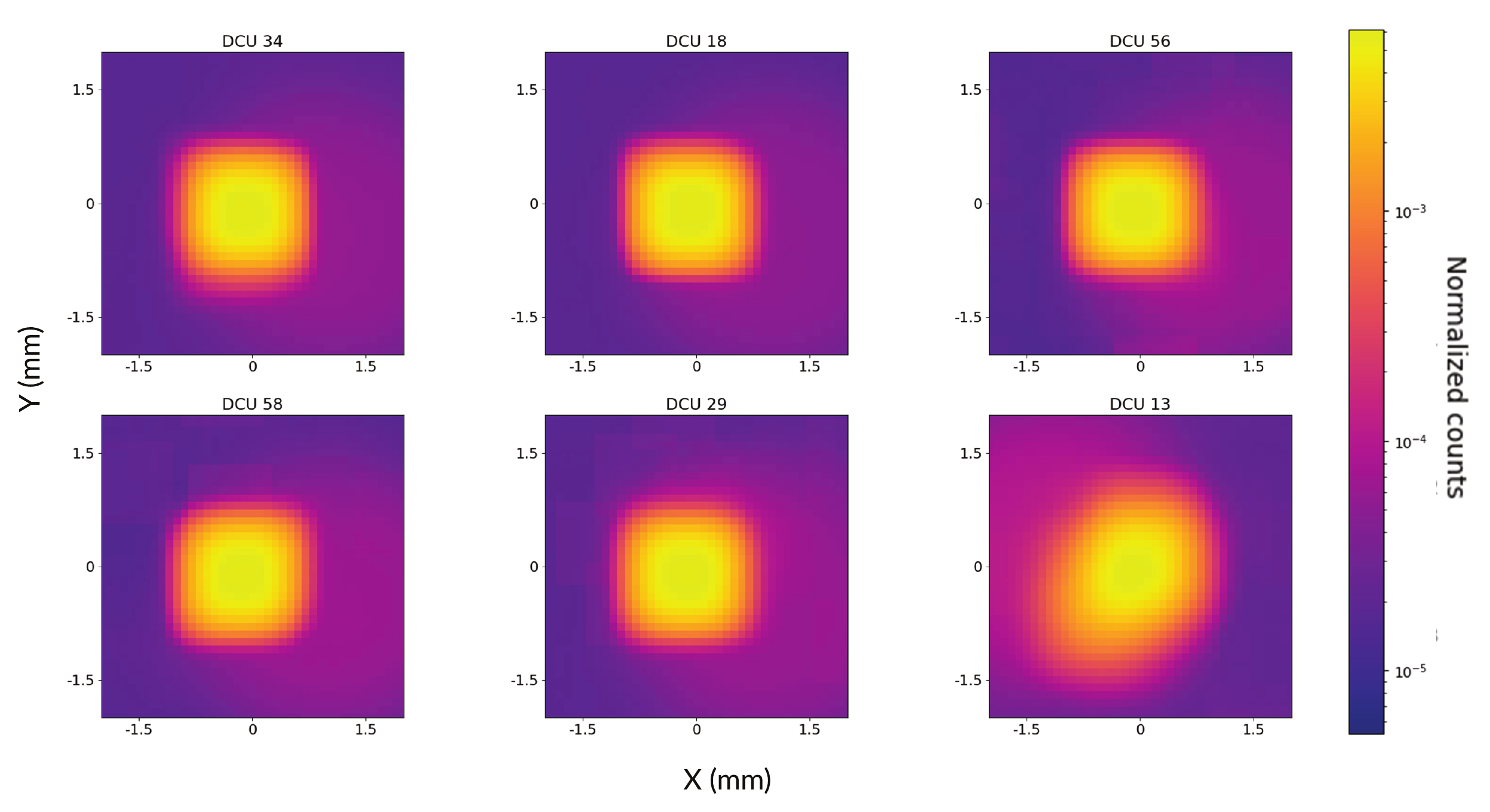} }}%
    \caption{`Reconstructed Stacked pixel image' of DCUs 34, 18, 56, 58, 29, and 13. The boundaries of a reconstructed stacked pixel image increase progressively from DCUs 34 to 13. The cross-correlated stacked pixel image of DCU 34 is well constrained within a $\simeq$ 15 $\times$ 15 sub-pixel grid, whereas DCU 13 requires at least a $\simeq$ 23 $\times$ 23 sub-pixel grid to constrain itself.}%
    \label{cross-corel}%
\end{figure}

A stacked DCU image as seen from the sky is reconstructed by cross-correlating (a convolution operation) a single P2 $\simeq 1.1$ mm mask element with a stacked DCU pixel obtained in section \ref{pix-stack}. The flux coming through from the mask element to the DCU pixel is assumed to be uniform, and hence accordingly distributed among all sub-mask elements and normalized to match the flux in a DCU stacked pixel. Figure \ref{cross-corel} shows stacked-pixel reconstructed images from DCUs 34, 18, 56, 58, 29, and 13, while figure \ref{cc_ideal}(right) shows a `reconstructed stacked pixel image' for an ideal DCU pixel. The `reconstructed stacked pixel image' is analogous to the point spread function of a DCU calculated at a sub-pixel resolution for a continuum X-ray flux between $\rm \simeq$ 5 - 25 keV\cite{basak_2020}.
When figures \ref{stack} and \ref{cross-corel} are set against each other, we can observe that the size of the reconstructed pixels increases from DCUs 34 to 13, and the shape of each conforms with that of the stacked pixel. The artifacts around each stacked pixel which lie beyond the mask element size, are smoothed out in the case of a reconstructed stacked pixel as seen in figure \ref{cross-corel}. In order to further analyze the reconstructed pixel images and quantify their morphologies, we select DCUs 34 and 13 as mentioned in section \ref{analysis}.

\begin{figure}%
    \centering
    \subfloat{{\includegraphics[width=0.98\textwidth]{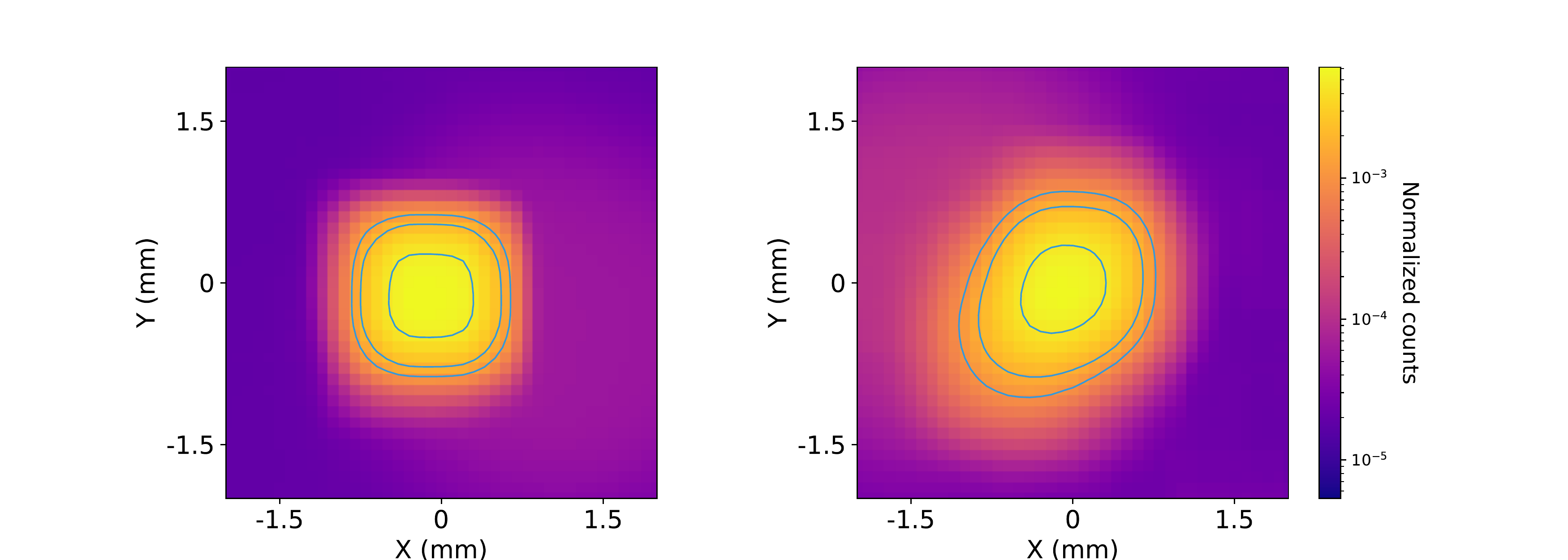}}}%
    \caption{Reconstruction of the detectors' response on the sky using the `stacked pixel' images for DCUs 34(left) and 13(right). The contour lines represent boundaries containing 50\%, 90\%, and 95\% of counts as one moves successively from the center to the outer confines of the reconstructed stacked pixel.}%
    \label{cc_zoom}%
\end{figure}

Figure \ref{cc_zoom} refers to zoomed-in reconstructed stacked pixel images from DCUs 34 and 13. The contour lines form boundaries within which 50\%, 90\%, and 95\% of events are recorded. DCU 34 represents a best-case scenario in P2 where 95\% of the total counts are constrained within 15 sub-pixels, whereas for DCU 13 the confines get extended up to 19-20 sub-pixels for the same fraction of counts. Moreover, the shapes of the contours in DCU 34 conform more closely to that of an ideal .6 mm $\times$ .6 mm sub-pixel grid, whereas the ones in DCU 13 are elongated along one of the diagonals matching the distribution of counts in the stacked pixel image.

\subsection{From stacked pixel to angular offsets}
\label{angres_sec}
The reconstructed stacked pixel images, along with their contour boundaries are used to estimate the effective pixel sizes in each DCU. The presence of variations in DCU stacked pixels relative to the mask hole pattern affects the angular resolution in a coded-aperture telescope\cite{Shen_2008}. The angular resolution for a DCU in a P2-like coded aperture telescope is given by\cite{caroli87}:

\begin{equation}
\label{ang_res}
    \Delta_{\rm ang} = \rm tan^{-1} \bigg(\frac{\sqrt{S_m^2 + S_d^2}}{f}\bigg) \hspace{0.5em},
\end{equation}

\noindent where $\Delta_{\rm ang}$ is the angular resolution, $\rm S_m$, and $\rm S_d$ are the sizes of the mask-hole and the pixel respectively, $\rm f$ is the focal length of the telescope, which is essentially the distance between the mask and the detector plane. For a P2-like telescope, $\rm S_m$ is 1.1mm and $\rm f$ is 888 mm. To calculate $\rm S_d$, we use figure \mbox{\ref{cc_ideal}} (left) i.e. an ideal stacked pixel and $\rm S_m$ to be 1.1 mm. The term on the numerator in equation \mbox{\ref{ang_res}} turns out to be $\rm \sqrt{S_m^2 + S_d^2} \simeq 1.26 mm$. We draw contour boundaries at 1.26mm on the ideal reconstructed stacked pixel and calculate the fraction of counts encompassed within ($\simeq 95 \%$). The 95\% contour boundaries as inferred from figure \mbox{\ref{cc_zoom}}, can then be used to set the effective pixel sizes ($\rm S_d$) to $\simeq$ 1.4 mm and $\simeq$ 1.9 mm for DCUs 34 and 13 respectively. Since $\rm S_d$ values are an approximation, we henceforth refer to the term defined by `$\Delta_{\rm ang}$' as angular offsets, a proxy for angular resolution. 


If a hypothetical P2-like telescope is primarily composed of DCU 34 like DCUs, its angular offset is $\rm \simeq$ 6.7 arcmins, while if it is constructed using DCU 13 like DCUs, the angular offset turns out to be $\rm \simeq$ 8.5 arcmins. The design specifications of HREXI detector planes have not been finalized yet. Hence, is it imperative to look at the changes in the effective pixel sizes, and by extension to the angular offsets as a function of the mask size element $\rm S_m$. 

\begin{figure}%
    \centering
    \subfloat{{\includegraphics[width=0.98\textwidth]{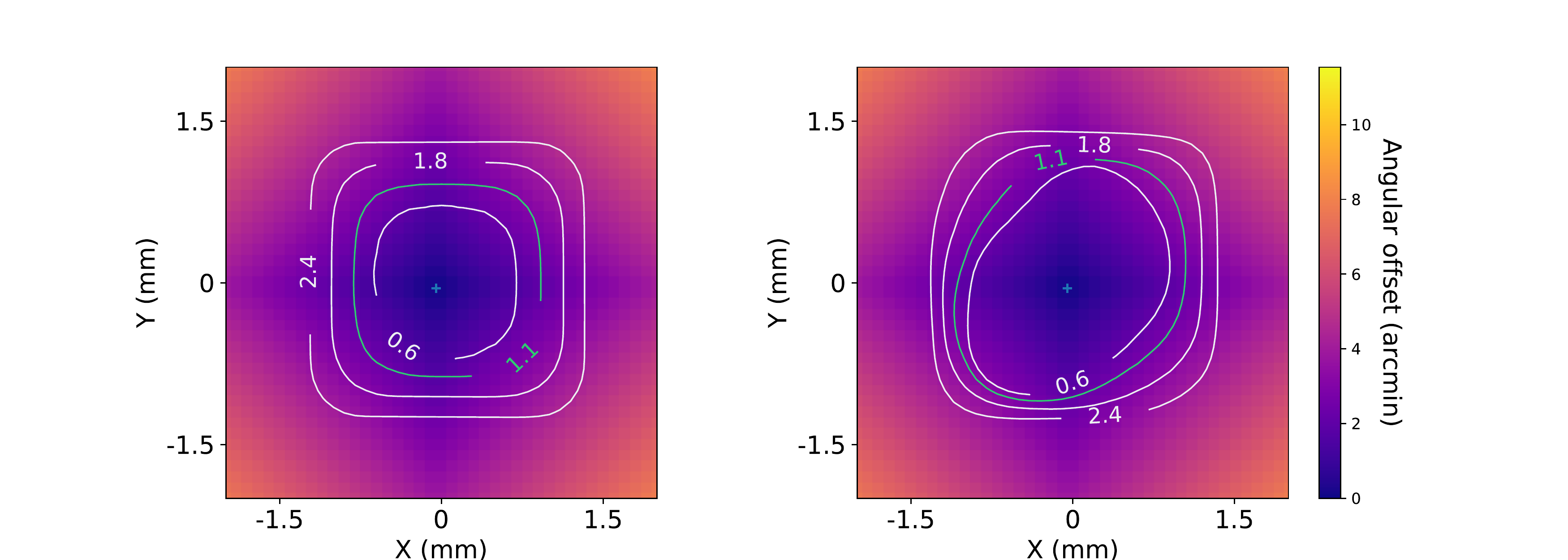} }}%
    \caption{Reconstructed stacked pixel images for DCUs 34(left) and 13(right) for different mask element sizes. The contour lines represent the effective pixel size ($\rm \simeq 95\%$ counts boundaries), while the labels illustrate the values of mask element size ($\rm S_m$) in mm. The color scale (s) illustrates the effective angular offset of the pixels from the pixel center, which in turn is represented by a `+'}. The blue contour line corresponding to 1.1 mm represents the mask element ($\rm S_m$) size of a P2-type telescope.%
    \label{cc_mask}%
\end{figure}

Figure \ref{cc_mask} demonstrates this for DCUs 34 and 13 by reconstructing stacked pixel images for different mask sizes. The contour lines represent the effective pixel size boundaries, while the labels illustrate the values of $\rm S_m$ in mm. The background(s) along with its color scheme can be used to get an estimate of the angular offsets of the pixels from their centers (annotated by a `+') for a fixed value of $\rm f = 888.0$  mm.  The presence of regular contours in the DCU-34 reconstructed image(left) implies a quasi-linear increase in the effective pixel size, with increasing values of the mask element size. DCU 13 has contours elongated along one of the diagonals for smaller values of $\rm S_m$. However, the shapes approach that of a regular mask pixel for larger values ($\simeq 2$ mm) i.e. when $\rm S_m$ tends towards $\rm S_d$.



The $\rm \Delta_{ang}$ along with signal-to-noise ratio (SNR) is used to determine the source localization of a coded-aperture telescope. The explicit calculation(s) of SNR is predicated upon the distribution of open and closed mask elements\cite{skinner08}, the background rate(s), the position of a pixel in a particular DCU, and more importantly the position of a DCU in the P2 detector plane. Pixels around the edge of a DCU are exposed to additional background radiation on account of the gaps introduced by the wire bonds to control and readout NuASICs\cite{hong2017}. Pixels adjacent to or in the vicinity of disabled hot pixels also have a proclivity to exposure to more background\cite{hong2017}. Moreover, in-flight calibration(s) have clearly shown that there is a gradual increase in background counts as one moves to the edge of a DCU\cite{hong2017} thereby making estimates of the SNR and by extension the angular resolution non-trivial. 

\section{Detector stacked pixel}
\label{detector_anal}
P2 detector comprises of an $\rm 8 \times 8$ array of DCUs and the HREXI detector plane(s) is slated to have an even bigger array of DCUs (at least $\rm 16 \times 16$)\cite{josh2019}. Hence, we construct detector `stacked pixel image(s)' averaged over all DCUs in the entire detector plane followed by the calculation of angular offset(s). Equation \ref{pixstack} can be modified such that it takes into account the mean contribution from each DCU in the detector plane,

\begin{equation}
\label{det-stack}
    \rm D_{i,j} = \frac{1}{N} \sum_{n = 1}^{N} P_{i,j|n} \hspace{0.5em},
\end{equation}

Here, $\rm D_{i,j}$ represents the counts in a stacked detector sub-pixel $\rm (i,j)$. The index $\rm n$ represents the DCU number in a $\rm N$ DCU detector plane. The other terms and indices in equation \ref{det-stack} are the same as equation \ref{pixstack}. Equation \ref{det-stack} can be modified to construct `stacked pixel image(s)' of a hypothetical HREXI detector plane comprising an ensemble of P2-type DCUs in terms of their weighted averages.

\begin{equation}
\label{det-weight}
       \rm  D_{i,j} = W_n \sum_{n = 1}^{N} P_{i,j|n} \hspace{0.5em},
\end{equation}

Here $\rm W_n$ denotes the weighted contribution from each DCU to the detector `stacked pixel image'. For a HREXI detector plane made up of 50\% DCU 34 like detector and 50\% DCU 13 like detectors, $\rm N = 2$, $\rm W_1 = 0.5 $ and $\rm W_2 = 0.5$. 

\begin{figure}%
    \centering
    \subfloat{{\includegraphics[width=0.98\textwidth]{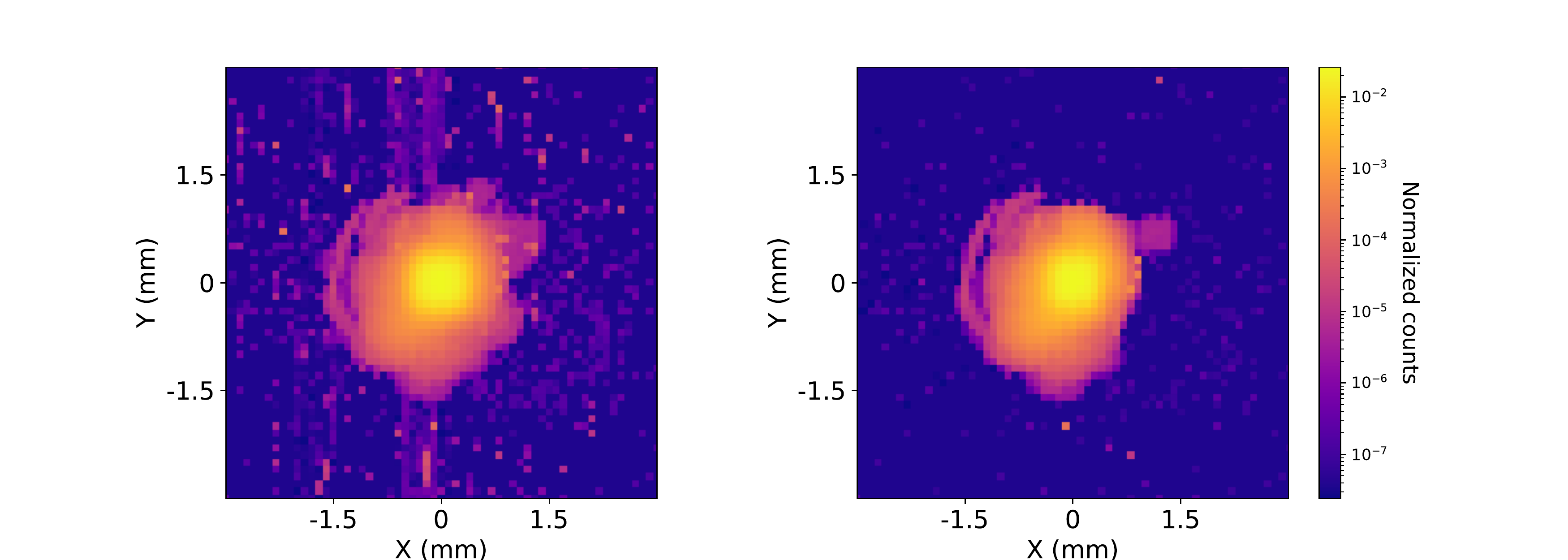} }}%
    \caption{Left: Stacked pixel image for the entire P2 detector plane. Right: Stacked pixel image for hypothetical HREXI detector plane made from a 50:50 split of DCU 34 and DCU 13-like detectors.}%
    \label{P2_stack}%
\end{figure}

\begin{figure}%
    \centering
    \subfloat{{\includegraphics[width=0.99\textwidth]{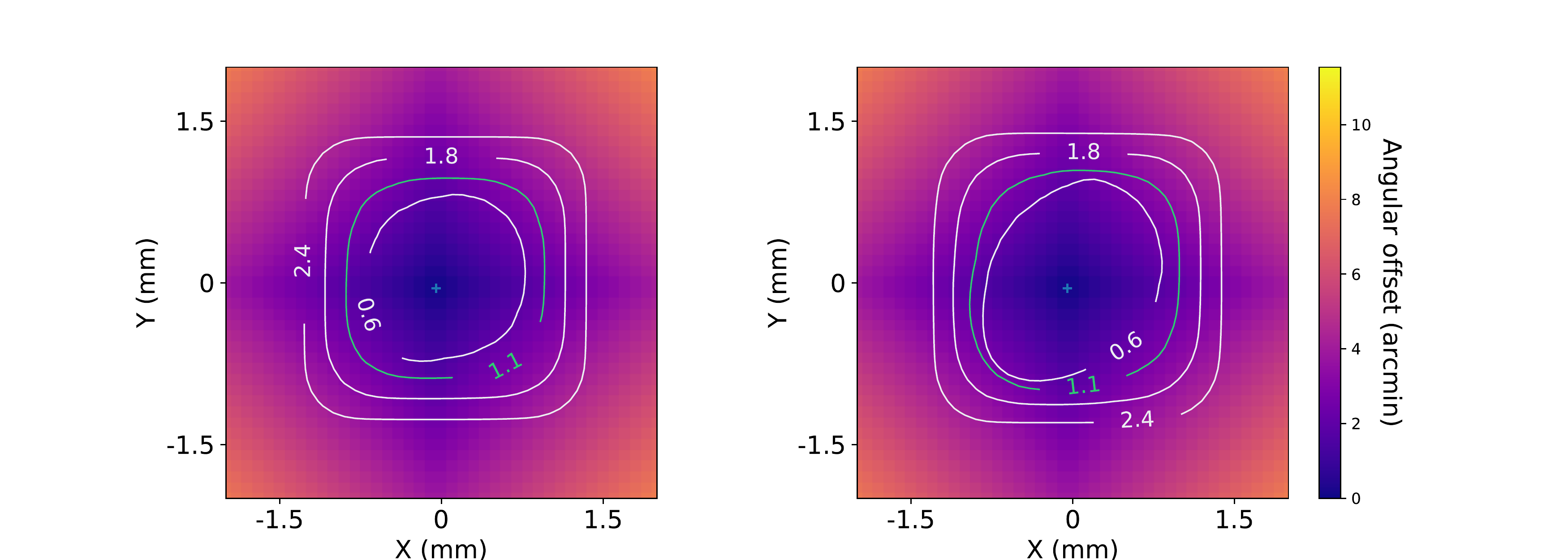} }}%
    \caption{Reconstructed stacked pixel images for P2 and a hypothetical HREXI detector plane for different mask element sizes. The contour lines represent the effective pixel size ($\rm \simeq 95\%$ counts boundaries), while the labels illustrate the values of mask element size ($\rm S_m$) in mm. The colorbar(s) illustrates the effective angular offset(s) of the pixels from the center, which in turn is represented by a `+'. The blue contour line corresponding to 1.1 mm represents the mask element size ($\rm S_m$) of a P2 type telescope.}%
    \label{cc_hx_det}%
\end{figure}

Figure \ref{P2_stack} shows the stacked pixel image for a P2 detector plane, along with a sample HREXI detector plane constituted from a 50:50 split of DCU 34 and 13-like detectors. The presence of disabled hot pixels, the removal of elastomeric bond separation in DCUs along with other distortions in the entire plane result in a noisier image in the case of P2. Cross-correlating a detector stacked pixel with different sizes of mask elements, gives us an estimate of the detector's effective pixel size, and in turn the detector's angular offset for a fixed focal length $\rm f$ without any statistical considerations. The contour lines in figure \ref{cc_hx_det} represent the effective pixel sizes both in terms of sub-pixel units and arcmins for different values of $\rm S_m$, while the blue line demonstrates the case for P2 type detector i.e. $\rm S_m = 1.1$mm.

The primary parameter of interest from the reconstructed stacked pixel image i.e. the effective pixel size turns out to be $\simeq$ 18 sub-pixels for a P2 type detector, whereas it is $\simeq$ 19 sub-pixels for an HREXI detector plane comprising of an equal number of DCU 13 and DCU 34 type detectors. As mentioned in section \mbox{\ref{angres_sec}}, the design specifications of the next generation HREXI detector plane(s), especially in terms of $\rm S_m$ and $\rm f$ have not been finalized. Figures \ref{cc_mask} and \ref{cc_hx_det} already illustrate the angular offset variations for different values of $\rm S_m$, and also that the effective pixel sizes tend towards the mask element size for sufficiently large values of $\rm S_m$, further illustrating the crucial interplay between $\rm S_m$ and $\rm S_d$ to optimize source localization\cite{skinner08, caroli87}.

\section{Correcting detector distortions - the centroid approach}
\label{correction}
\begin{figure}%
    \centering
    \subfloat{{\includegraphics[width=0.99\textwidth]{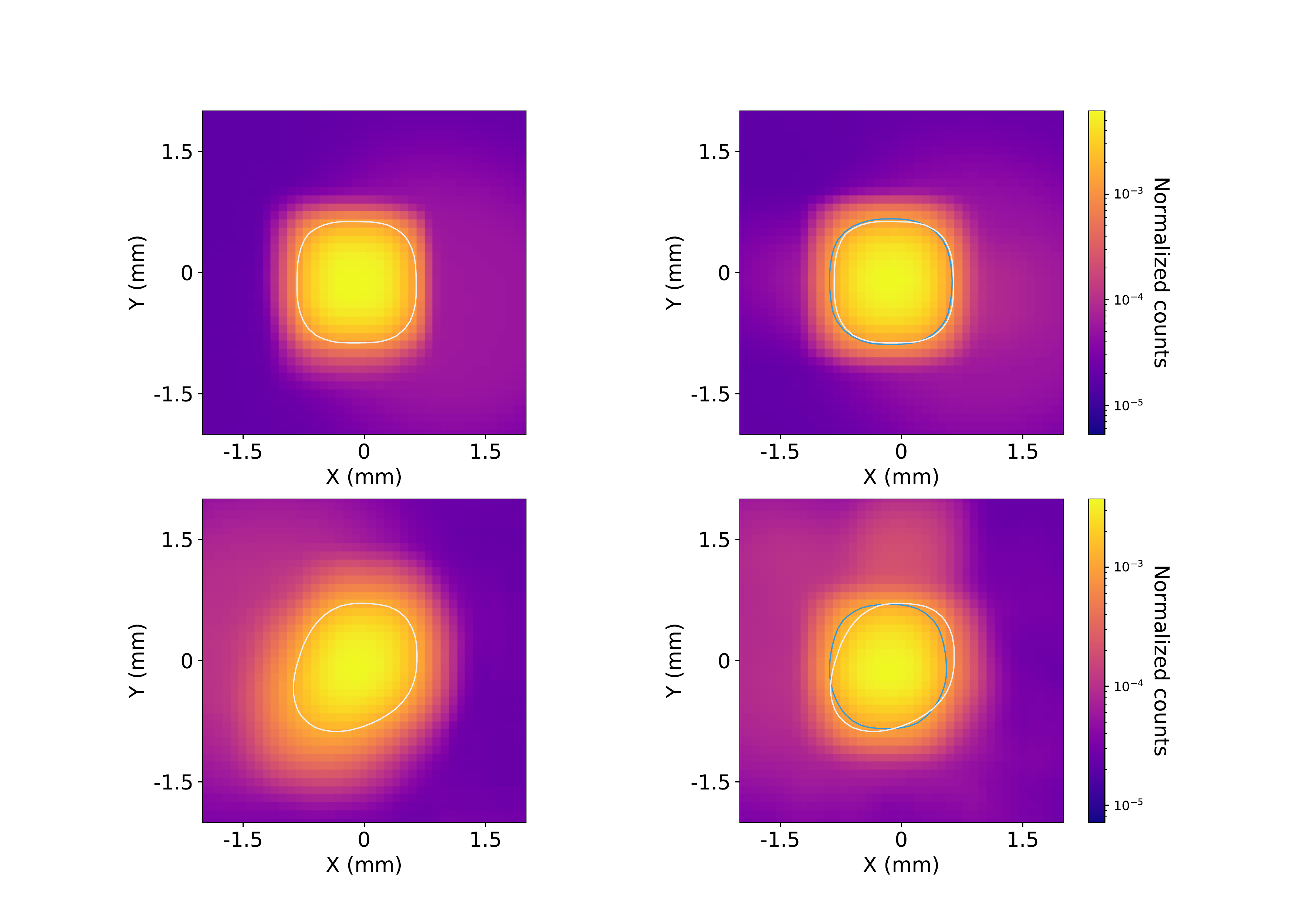} }}%
    \caption{ Top: `Reconstructed stacked pixel images' for DCUs 34 before(left) and after(right) the centroid corrections have been applied. Bottom: `Reconstructed stacked pixel images' for DCUs 13 before(left) and after(right) the centroid corrections. The white and blue contours(s) encompass the effective pixel boundaries ($\rm S_d$) before and after the application of the centroid corrections for each DCU.}%
    \label{33_centroid_stack}%
\end{figure}

A method of mitigating asymmetrical stacked pixel shapes would be to superimpose pixels on top of each other after aligning their centroid(s) rather than embossing them on the ideal $6 \times 6$ sub-pixel grid. Figure \ref{33_centroid_stack} (top) elucidates the reconstructed stacked pixel image along with its effective pixel size for DCU 33, before and after the centroid corrections have been applied, while figure \ref{33_centroid_stack}(bottom) does the same for DCU 13. DCU 34 looks very similar to that of figure \ref{cc_zoom} whereas the asymmetric elongation along one of the diagonals for DCU 13 is mitigated. This essentially implies that the non-uniform charge diffusion occurrences beyond pixel boundaries can be corrected by realigning pixels, whereas such events occurring uniformly across all four-pixel boundaries are harder to account for. The effective pixel size for DCU 13 now reduces to $\simeq$ 18 sub-pixels while there is virtually no change for DCU 34.

\begin{figure}%
    \centering
    \subfloat{{\includegraphics[width=0.90\textwidth]{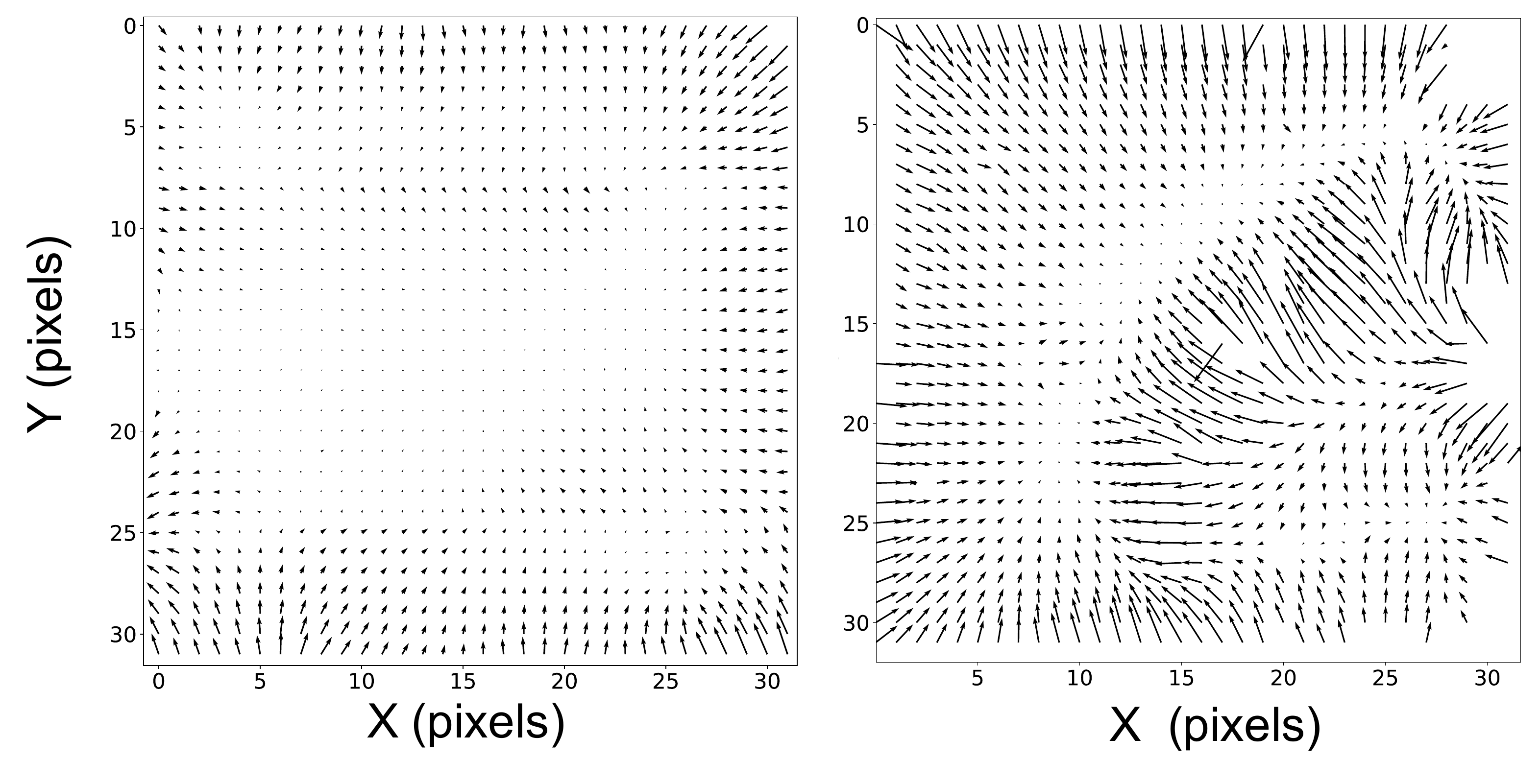} }}%
    \caption{Vector maps for aligning each pixel of DCUs 34(left) and DCU 13(right) from its true center to its calculated centroid. The X-Y axes have been labeled in pixels, as each vector in the map represents the correction per DCU pixel. The size of each pixel is $\rm \simeq 0.604$ mm as mentioned in section \mbox{\ref{intro}}. }%
    \label{vector_map}%
\end{figure}

Figure \ref{vector_map} illustrates vector maps for DCUs 34 and 13, aligning each pixel from its true center to its calculated centroid. The length and the distribution of vectors are measures of non-uniformities present in a DCU region. Shorter and non-existent vectors imply uniform regions whereas longer vectors flock together in an imperfect region of a DCU. These vector maps calculated explicitly for each DCU should be used in tandem with in-flight observations to correct for systematic pixel position uncertainties before any contribution from the background. 

\begin{figure}%
    \centering
    \subfloat{{\includegraphics[width=0.99\textwidth]{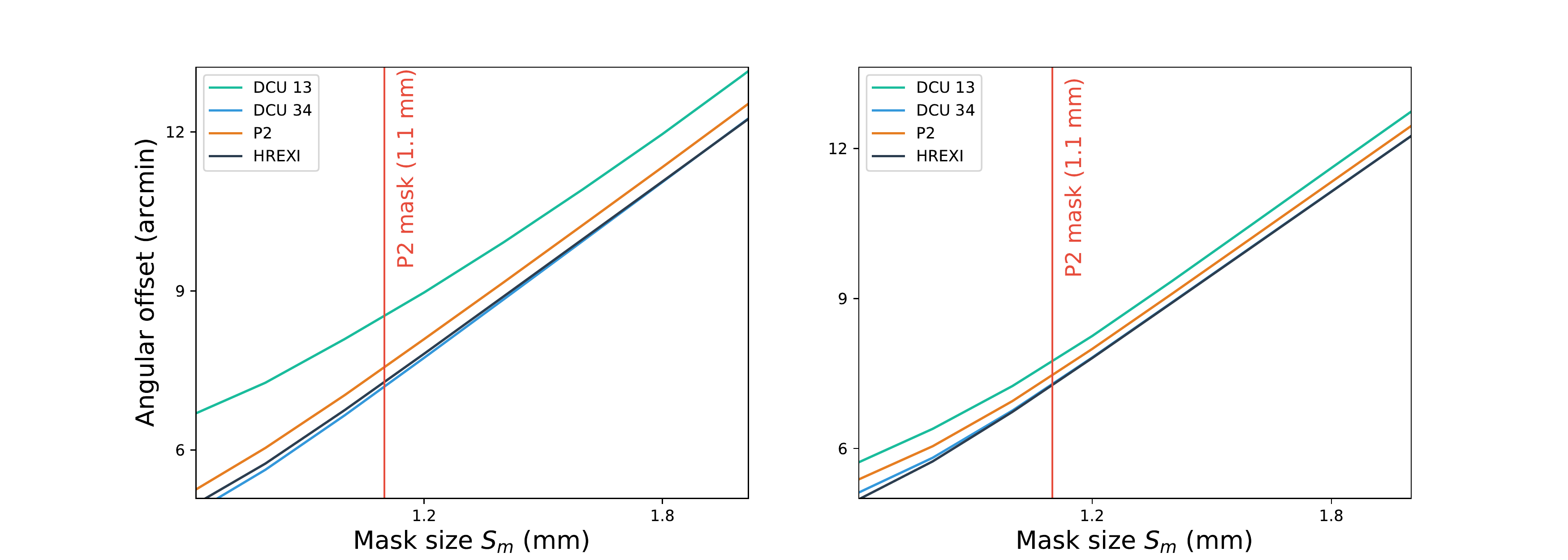} }}%
    \caption{The variation(s) of angular offset(s) for different mask size elements ($\rm S_m$) before (left) and after (right) application of the centroid correction routine. The red line corresponds to the P2 mask size element of 1.1 mm.}%
    \label{offset}%
\end{figure}

\begin{figure}%
    \centering
    \subfloat{{\includegraphics[width=0.99\textwidth]{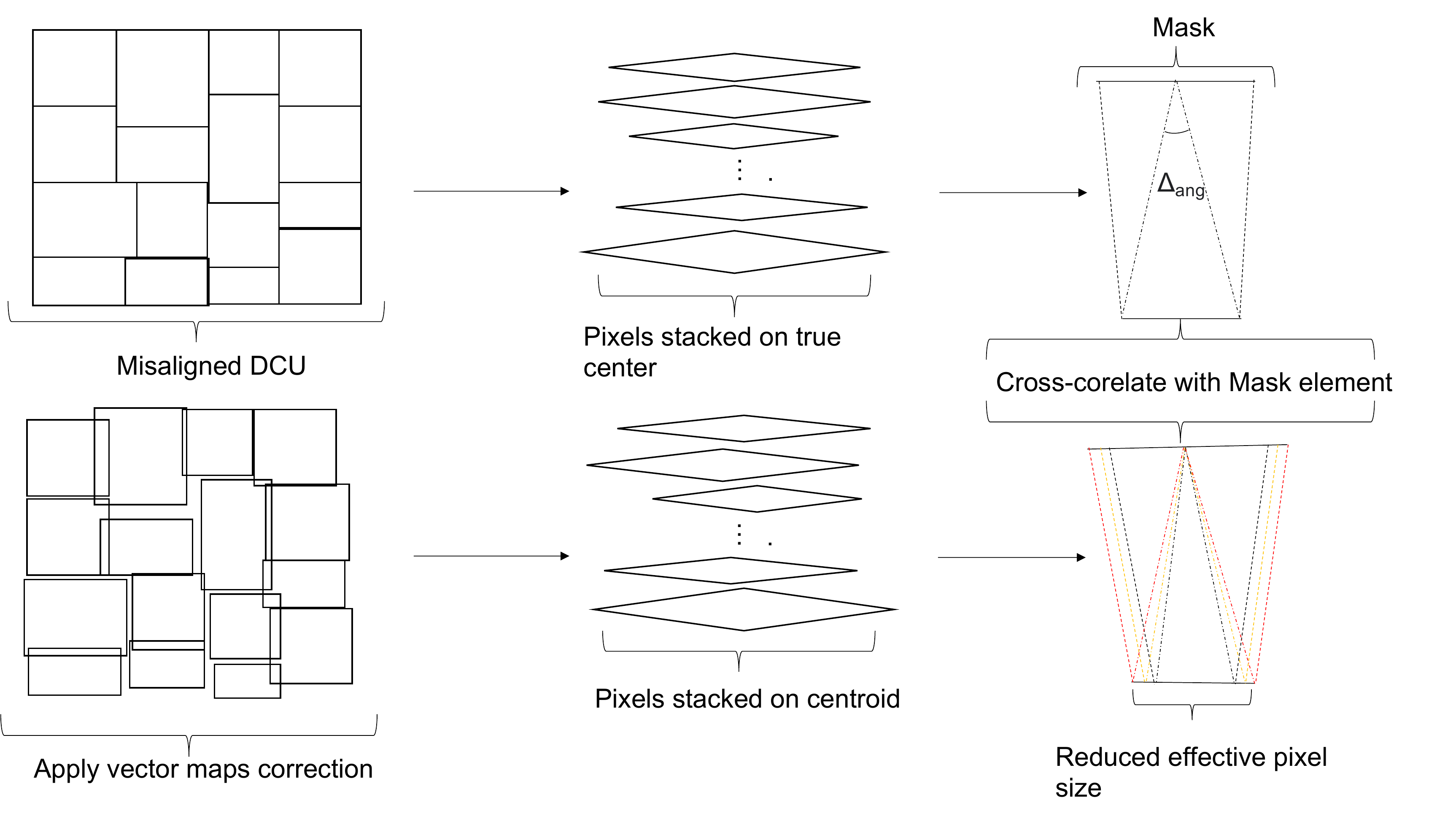} }}%
    \caption{Schematic diagram showing a DCU before (Top) and after (bottom) applying vector maps correction, pixel stacking, and then cross-correlation with a mask element, with an annotated $\rm \Delta_{ang}$. The colored lines on the bottom right diagram signify different sizes of mask elements used and their corresponding $\rm \Delta_{ang}$s.}%
    \label{cc-schematic}%
\end{figure}

The change in the contour boundaries before and after application of the centroid correction(s), and by extension of the angular offset is explicitly illustrated in Figure\ref{offset} for DCUs 13 and 34, the entire P2 detector and a hypothetical HREXI detector comprising of 50\% DCU 13 and 50\% DCU 34 type detectors. The reduction in the angular offset becomes more apparent for DCU 13 in figure \ref{offset} (right), whereas things remain virtually identical for the other three cases. 

 Figure \ref{cc-schematic} illustrates a schematic diagram of a DCU before (Top) and after (bottom) applying vector maps correction, pixel stacking, and then cross-correlation with a mask element. The colored lines on the bottom right diagram signify different sizes of mask elements and the resultant stacked pixel from an edge-on view, along with their corresponding angular offsets $\rm \Delta_{ang}$. The change in the effective pixel boundaries(s) before and after the correction(s) can be used to estimate the change in the relative source counts within a reconstructed stacked pixel. This translates to a change in the SNR for a given DCU, as events outside the pixel boundaries(s) would be incorrectly assigned to the neighboring pixel(s) while applying the cross-correlation algorithm and lead to a degradation of the source signal. The source detection SNR ($\rm  \hat{S}/\sigma_S$) for a coded aperture telescope can be quantified from section 3 in Skinner et. al. '08\mbox{\cite{skinner08}} as a function of source counts and background counts ($\rm C_S, C_B$), detector area (A), and coding fraction (F)


\begin{equation}
   \rm  \hat{S}/\sigma_S = C_S \sqrt{\frac{ F A (1 - F)}{F C_S  + C_B}} \hspace{0.5em},
    \label{snr}
\end{equation}

A relative gain in source counts $\rm (dC_S)$, leads to a corresponding change in the background counts $\rm -dC_B$, and hence the relative change in SNR can be quantified as a function of the relative change in source counts $\rm (dC_S/C_S)$, and a background scaling parameter $\rm \alpha = C_B/C_S$

\begin{equation}
   \rm  \frac{d (\hat{S}/\sigma_S)}{\hat{S}/\sigma_S} = \frac{d C_S}{C_S} \bigg[ 1 - \frac{F}{2 C_S (F + \alpha)}\bigg] \hspace{0.5em},
    \label{snr-2}
\end{equation}

Here, $\rm dC_S = C_{S} - C_{S'}$. $\rm C_{S}$ and $\rm C_{S'}$ refer to the source counts in the stacked pixel before and after the centroid corrections. For $\rm n^{th}$ DCU in the detector plane, $\rm C_{S} = \sum P_{n}$, whereas $\rm C_{S'} =  \sum P^{'}_{n}$. The summation for both cases should be along the contour boundary $\rm C'$ after the centroid corrections. $\rm P_{\rm n}$ and $\rm P^{'}_{\rm n}$ illustrate the DCU/detector stacked sub-pixel before and after the centroid corrections. The coding fraction (F) refers to the ratio of open to closed mask elements and is fixed at 0.5. For DCU 34 the relative gain in SNR is almost negligible $\rm ( \leq 1 \%)$ since the reconstructed stacked pixel image looks identical before and after the centroid corrections, while it is about $\rm \simeq 4 \%$ for DCU 13 in a P2 type telescope, without additional background. The relative gain in SNR becomes more pronounced as one transitions to smaller mask hole sizes since the DCU non-uniformities have bigger effects on the contour boundaries as illustrated in figure \ref{cc_mask}. The $\Delta_{\rm SNR}$ can be as high as $8 \%$ for DCU 13 and a mask hole size of 0.6 mm. The primary results from this set of analyses have been summarized in table \ref{parameters} for $\rm S_m = 1.1 mm$ and $\rm f = 888 mm$ i.e. the P2 telescope parameters.

\begin{table}
\begin{center}
\caption{Primary results from four sets of detector planes P2, HREXI, and the two DCUs 13 and 34, for a zero external background case.}
\begin{tabular}{||m{4em} | m{4em} | m{5em} | m{6 em} | m{7em} | m{3em}||}

 \hline
Detector & Pixel size (mm) & Offset (arcmin) & Pixel size after centroid (mm) & Offset after centroid (arcmin) & $\Delta_{\rm SNR}$ \\ [0.5ex] 
 \hline\hline
 DCU 13 & 1.91 & 8.53 & 1.88 & 7.6 &  4\%  \\ 
 \hline
 DCU 34 & 1.43 & 7.13 & 1.58 & 7.1 & $\leq$1\% \\ [1ex] 
 \hline
P2 & 1.87 & 7.6 & 1.84 & 7.53 & $\simeq$ 1\% \\ [1ex] 
 \hline
HREXI & 1.82 & 7.3 & 1.81 & 7.25 & $\leq$1\% \\ [1ex] 
 \hline
\end{tabular}
\label{parameters}
\end{center}
\end{table}

\section{Discussions and Future work.}
\label{conclusions}

The efficiency of an X-Ray telescope to resolve different sources in the sky depends directly on the capability of the detector to correctly reconstruct event positions on the detector plane.  In order to characterize the deviation of our detector systems from our ideal 32~$\times$~32 anode pattern, we have stacked the individual pixel responses relative to the ideal anode pattern using equation \ref{pixstack} which calculates the effective average pixel response and morphology of each DCU.  Expanding this further, in section \ref{detector_anal} we have characterized the average pixel size and shape for the entire P2/HREXI detector plane, and used these quantities to probe the effects of our imperfect detectors on angular offset(s) for a number of different scenarios (see section \ref{ang_res}). Section \ref{correction} illustrates a strategy to eliminate some of the non-uniformities using pixel centroiding. Allen et. al. \cite{subpix} calculates DCU vector maps explicitly for the next generation of HREXI detectors. Moreover, there have been attempts to develop a detector position reconstruction algorithm as a function of interaction depth in the CZT crystal and the $\rm V_{bias}$ between the anode and the cathode\cite{subpix, josh2019_2}. Studies of pixel sizes as a function of $\rm V_{bias}$ phenomenologically reveal the decrease in size of these distortions without a change in their morphology and pattern\cite{subpix} as $\rm V_{bias}$ increases. This can be qualitatively attributed to the decrease in charge diffusion in direction(s) orthogonal to the direction of photon propagation with the increase in $\rm V_{bias}$. At present, the exact physical origins of these anomalies are unknown and require further work. Future investigation is planned to investigate the origin of these anomalies by extracting a cross-section of the crystal to enable direct characterization of the CZT bulk properties. Additionally, a comparison with 3mm detectors is also planned, which would delve deeper into the quantification of pixel sizes and morphologies as a function of interaction depth.

The sensitivity of a hard X-ray telescope is also dependent on the contribution of the stochastic background rate to the SNR, hence co-simulating response from the HREXI detector plane along with an energy-dependent background model is essential to get a more accurate estimate of the background rate. Since the final design specifications of the next generation detector planes have not been zeroed in on yet, and the background rates are heavily dependent on the geometry, orientation, and the material of shields used around the detector plane\cite{spie11}, this analysis beyond the scope of this current article. However, a zeroth order calculation of SNR has been included in section \ref{correction}.

HCF's capability to scan entire large-area detector planes on timescales of $\simeq$~3 days\cite{basak_2020} enables rapid characterization of each of the individual detectors' impact on the angular resolution of the full detector plane.  In future works, imaging simulations using detectors with non-ideal spatial responses, that mimic those exemplars thus far observed for 5~mm and 3~mm CZT material, will be carried out to inform selection criteria for future CZT-based telescopes.  Furthermore, updated characterizations of CZT quality from recently fabricated detectors will be integrated into this analysis in order to perform a trade study for the evaluation of the cost-benefit ratio of detector screening.  Ultimately this trade study would be carried out and applied to the fabrication and integration of future hard X-ray telesocpes utilizing the code-aperture imaging technique.

\section{Acknowledgements}
This work was supported by NASA grant NNX17AE62G. The authors would like to thank Stan G. Corteau and Mike Mckenna of Harvard John A. Paulson School of Engineering and Applied Sciences machine shop for their support during the fabrication of HCF. The authors would also like to thank the anonymous reviewer(s) for their comments which improve the quality, readability, and content of the paper.

\bibliography{main_1}   
\bibliographystyle{spiejour}   

\vspace{1ex}
\noindent Biographies and photographs of the authors are not available.

\listoffigures

\end{spacing}
\end{document}